\documentclass[a4paper,10pt,conference]{ieeeconf}
\IEEEoverridecommandlockouts
\usepackage{amsmath,amssymb}
\usepackage{graphicx}
\usepackage{booktabs}
\usepackage{cite}
\usepackage{xcolor}

\newtheorem{theorem}{Theorem}
\newtheorem{lemma}[theorem]{Lemma}
\newtheorem{proposition}[theorem]{Proposition}

\newtheorem{definition}[theorem]{Definition}
\newtheorem{remark}[theorem]{Remark}

\title{\LARGE\bf AFRA: Age-Threshold Frame-Based Random Access for
Timely Communication in Multi-Agent Systems}
\author{Jiedan Tan and Fang Liu%
\thanks{J. Tan and F. Liu are with the College of Electronics and Information
Engineering, Shenzhen University, Shenzhen, China, e-mails:
2510044003@mails.szu.edu.cn; liuf@szu.edu.cn.}%
\thanks{The corresponding author is Fang Liu.}}

\begin{document}
\maketitle
\thispagestyle{empty}
\pagestyle{empty}
% Keep natural vertical spacing instead of stretching short columns.
\raggedbottom

\begin{abstract}
Timely state updates are essential for coordination and monitoring in
multi-agent systems, such as teams of mobile robots exchanging state
information. When agents share a wireless channel, collisions and
slow convergence of distributed access schemes can degrade information
freshness over finite operating periods. This paper proposes age-threshold
frame-based random access (AFRA), a distributed medium access control
protocol for minimizing age of information (AoI). Once its age reaches a
threshold, an agent starts a local frame with a mandatory transmission and,
if unsuccessful, makes one further attempt at a uniformly selected remaining
slot. Each attempt carries a freshly generated update. AFRA requires only
local state and acknowledgment feedback, with shared slot boundaries but
independently timed local frames. We prove almost-sure convergence from any
legal initial state to collision-free periodic operation when the frame
length $L$ and age threshold $T$ satisfy $L\ge T>N\ge2$, where $N$ is the
number of agents. We also derive closed-form expressions for steady-state
throughput, average AoI, and average peak AoI. Simulations demonstrate that
AFRA combines rapid convergence at small age thresholds with lower
window-averaged AoI than the existing age-aware random access schemes
evaluated in this study.
\end{abstract}


\section{Introduction}
Timely wireless communication supports coordinated planning and monitoring
in multi-agent systems. In robotic applications, communication constraints affect team
coordination~\cite{shi2021communication}, robot positioning for communication
coverage~\cite{pongsirijinda2023gmc}, and the transfer of collected data
to a data center~\cite{guo2018gathering}. For example, mobile robots report their positions,
battery levels, and task progress to a common access point (AP) to support
navigation and task allocation. Timely updates on exploration progress and
fleet status are also essential for human supervision~\cite{tian2024ihero}.
Such information loses value as it becomes
outdated. Age of information (AoI), defined as the time elapsed since the
generation of the latest successfully received update, provides an important
measure of information timeliness. When agents share a wireless channel,
their medium access control (MAC) scheme determines when updates can be
delivered and how often transmissions collide. Designing distributed MAC
schemes is therefore crucial to AoI optimization in multi-agent networks
without centralized transmission scheduling.

A broad range of MAC schemes has been studied for AoI optimization.
Age-unaware schemes optimize access parameters or schedules without basing
individual transmission decisions on instantaneous AoI. Slotted ALOHA offers
simple distributed access, but collisions and random waiting times limit its
AoI performance~\cite{yates2017status}. Munari~\cite{munari2021modern}
showed that irregular repetition slotted ALOHA (IRSA) can improve information
freshness through packet repetition and successive interference cancellation.
These gains require additional packet replicas and receiver processing,
while the frame duration introduces a tradeoff between decoding reliability
and update delay. Yue et al.~\cite{yue2023frame} analyzed frame slotted ALOHA,
where each updating source chooses one transmission slot per frame, showing
reductions in average AoI and its variability in dense networks. However,
performance remains sensitive to the frame length and interference, requiring
appropriate parameter tuning. Maatouk et al.~\cite{maatouk2020csma}
optimized CSMA backoff parameters to reduce AoI, although this approach
requires channel sensing. Liu et al.~\cite{liu2026sequence} developed an
AoI analysis and parameter optimization methodology for sequence-based
scheduling, enabling low-overhead operation without channel sensing or
receiver feedback. However, the resulting fixed transmission sequences
cannot adapt online to the current AoI.

Age-aware schemes instead use
freshness to regulate contention. Tripathi et al.~\cite{tripathi2023fresh}
proposed Fresh-CSMA, which uses age-dependent backoff to approximate
centralized max-weight scheduling; its practical implementation must balance
collisions against carrier-sensing and backoff overhead. Chen et al.~\cite{chen2022age}
proposed decentralized age-based thinning policies, including stationary
age-based thinning (SAT), which suppress transmissions with small potential
age reductions. Threshold slotted ALOHA (TSA), analyzed by
Yavascan and Uysal~\cite{yavascan2021threshold}, keeps recently updated
nodes silent and activates probabilistic access when their age reaches a
threshold. Although age-based thinning and TSA reduce contention and improve AoI,
eligible nodes still compete through random access, leaving collisions and
idle slots in steady operation. Ahmetoglu et al.~\cite{ahmetoglu2022mista}
proposed MiSTA, which improves on TSA by using a contention minislot and
probabilistic backoff to reduce data-slot collisions, but requires additional
control signaling and collision sensing. Dester et al.~\cite{dester2026onepersistent}
introduced 1-persistent TSA, which transmits deterministically at the
threshold and probabilistically thereafter. Under generate-at-will traffic,
it can achieve collision-free periodic operation, but convergence can be
prohibitively slow when the threshold approaches the number of nodes,
limiting its benefit over finite observation windows. These results motivate a MAC design
that combines low steady-state AoI with rapid convergence using only
acknowledgment feedback, without carrier sensing or additional contention
minislots.

To this end, we propose age-threshold frame-based random access (AFRA), a
distributed MAC scheme for timely multi-agent communication. When its age
reaches threshold $T$, an agent starts a local frame of length $L$ and
transmits at the frame head. If this attempt fails, it retries once at a
uniformly selected remaining slot. If both attempts fail, a new frame begins
after the current frame ends; a success returns the agent to threshold-based
waiting. This frame structure replaces independent per-slot random attempts
with a deterministic frame-head attempt and at most one random retry in each
local frame. Agents share slot boundaries, but their local frames
need not be aligned, and access decisions require only local state and
acknowledgment feedback.

Our contributions are threefold. First, we establish that AFRA converges
almost surely to a collision-free periodic regime from any legal initial
state when $L\ge T>N\ge2$, where $N$ is the number of agents. Second, we derive
closed-form expressions for steady-state throughput, average AoI, and average
peak AoI. Third, we evaluate transient behavior and finite-window AoI against
original TSA, 1-persistent TSA, and SAT, with TDMA as an ideal scheduling
reference. The simulations show that faster convergence allows AFRA to use
smaller age thresholds effectively, improving information freshness over
finite observation windows.

\section{System Model and Access Protocol}
\label{sec:system}

\subsection{Network and Channel Model}
\begin{figure}[!ht]
\centering
\includegraphics[width=\columnwidth]{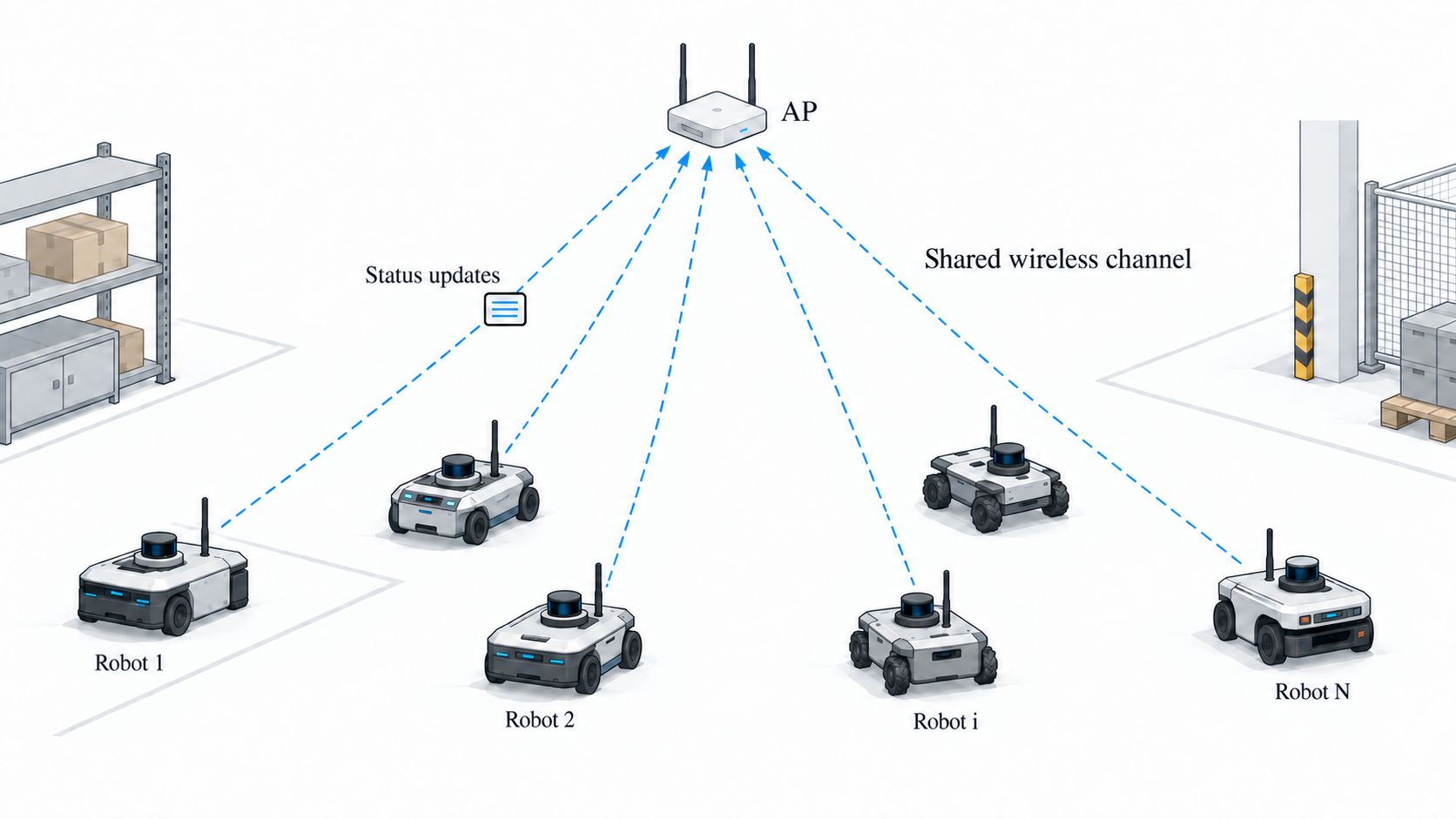}
\caption{Multi-robot status updating over a shared wireless channel to an
access point (AP). Dashed arrows indicate potential uplink communication
links, rather than simultaneous successful transmissions.}
\label{fig:robot-ap-scenario}
\end{figure}

As illustrated in Fig.~\ref{fig:robot-ap-scenario}, we consider $N\ge2$
agents that send state updates to a common access point (AP) over a shared
single-hop wireless channel. Each agent monitors an
independent process, such as a mobile robot's pose, battery level, or task progress. Time is
divided into slots indexed by $t\in\mathbb{Z}_{\ge0}$. All agents share slot
boundaries and decide whether to transmit at the beginning of each slot;
each report occupies one slot. There is no orthogonal channelization
and no central scheduler. Let $U_i(t)\in\{0,1\}$ indicate whether agent $i$
transmits in slot $t$, and let $Z(t)=\sum_{i=1}^{N}U_i(t)$ be the number of
transmitters. The channel is idle if $Z(t)=0$, delivers the report if $Z(t)=1$,
and experiences a collision if $Z(t)\ge2$, in which case all simultaneous transmissions fail.
We assume no capture effect, multi-packet reception, successive interference
cancellation, or transmission errors other than collisions. At the end of
each slot, error-free acknowledgment (ACK) feedback from the AP allows each
transmitting agent to determine whether its report was delivered before the
next slot begins. ACK feedback is the only receiver feedback, and agents
exchange no coordination messages.

\subsection{Age of Information}
Let $\Delta_i(t)$ be the age of agent $i$'s state at the beginning of slot $t$,
before the transmission decision, i.e., the number of slots elapsed since the
state most recently received by the AP was generated. Reports follow a
generate-at-will model: whenever an agent transmits, it samples its process
at the beginning of that slot and generates a fresh report;
otherwise no report is generated, and no queue of previously generated reports
is kept. Hence a retry after a collision is a new access attempt rather than a
retransmission of the collided report. Let
$Y_i(t)\triangleq U_i(t)\mathbf{1}\{Z(t)=1\}$ indicate a successful delivery
from agent $i$ in slot $t$. A successful delivery resets the age in the next
slot, while silence and collisions let it grow by one:
\begin{equation}
\Delta_i(t+1)=
\begin{cases}
1, & Y_i(t)=1,\\
\Delta_i(t)+1, & Y_i(t)=0.
\end{cases}
\label{eq:aoi-recursion}
\end{equation}
We evaluate age at slot beginnings throughout, including when defining
peak AoI below.

Let $H$ denote the observation window length in slots, covering
$t=0,\dots,H-1$. We use the following three long-run metrics, assuming
their limits exist. The time-average
AoI of agent $i$,
\begin{equation}
\overline{\mathrm{AoI}}_i
\triangleq
\lim_{H\to\infty}\frac{1}{H}\sum_{t=0}^{H-1}\Delta_i(t),
\label{eq:average-aoi}
\end{equation}
measures the typical freshness of its state at the AP over time. The average
peak AoI (PAoI) of agent $i$,
\begin{equation}
\overline{\mathrm{PAoI}}_i
\triangleq
\lim_{H\to\infty}
\frac{\sum_{t=0}^{H-1}\Delta_i(t)Y_i(t)}
{\sum_{t=0}^{H-1}Y_i(t)},
\label{eq:average-paoi}
\end{equation}
averages the slot-beginning age in successful transmission slots. These
values are the peaks of the sampled age trajectory before its resets.
This metric assumes infinitely many successful deliveries. Finally, the throughput
\begin{equation}
\eta
\triangleq
\lim_{H\to\infty}\frac{1}{H}\sum_{t=0}^{H-1}\sum_{i=1}^{N}Y_i(t)
\label{eq:throughput}
\end{equation}
is the average number of reports delivered per slot by the whole system, bounded
by $0\le\eta\le1$ since a slot delivers at most one report. Whenever a metric
is identical for all agents, the subscript $i$ is omitted.

\subsection{Distributed Access Protocol}
\begin{figure*}[t]
\centering
\includegraphics[width=0.95\textwidth]{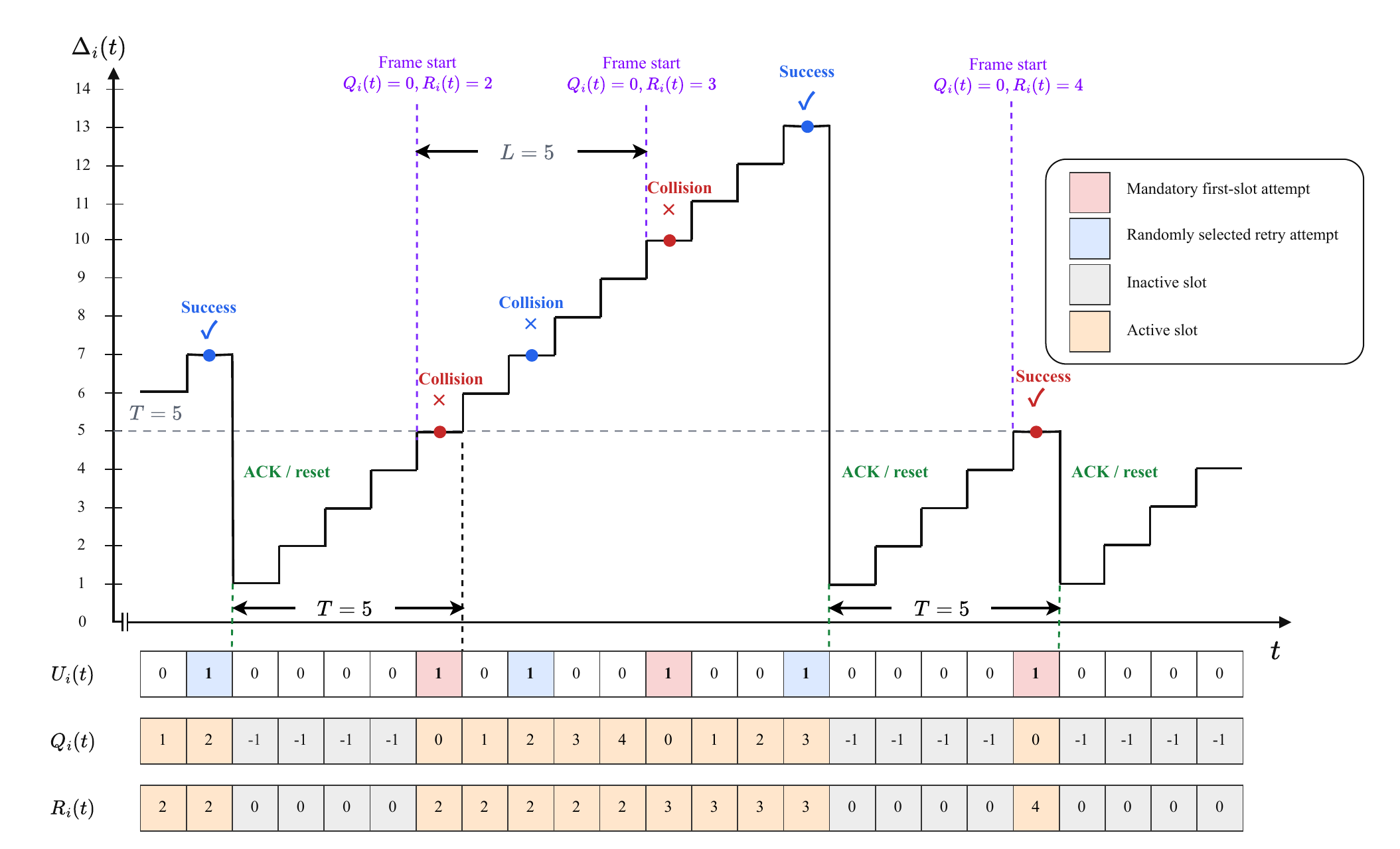}
\caption{Example timeline of the AFRA protocol for a single agent with
$T=L=5$. The AoI trajectory is aligned with the transmission decision $U_i(t)$,
frame position $Q_i(t)$, and retry position $R_i(t)$. A successful transmission
resets the AoI to one in the next slot and cancels any pending retry.}
\label{fig:protocol-timeline}
\end{figure*}
We propose the age-threshold frame-based random access (AFRA) protocol, an
asynchronous local-frame scheme with a positive integer age threshold $T$
and an integer frame length $L\ge2$, common to all agents. Besides
$\Delta_i(t)$, each agent maintains two local variables: the frame
position $Q_i(t)\in\{-1,0,\dots,L-1\}$, where $Q_i(t)=-1$ means inactive and
$Q_i(t)=q\ge0$ means active at position $q$ of the current frame (position $0$
is the frame head), and the retry position $R_i(t)\in\{0,\dots,L-1\}$.
The value $R_i(t)=0$ is reserved for the inactive state. At each frame start,
the agent independently draws a retry position uniformly from
$\{1,\dots,L-1\}$. The access rule is:
\begin{enumerate}
\item \emph{Inactive.} While $\Delta_i(t)<T$, agent $i$ keeps
$Q_i(t)=-1$, $R_i(t)=0$, and remains silent.
\item \emph{Frame start.} When $\Delta_i(t)\ge T$ and $Q_i(t)=-1$, the agent
starts a local frame: it sets $Q_i(t)\leftarrow0$, draws
$R_i(t)\sim\mathrm{Uniform}\{1,\dots,L-1\}$, and mandatorily transmits at the
frame head. In each subsequent slot without a success, the frame position
advances by one until the frame ends.
\item \emph{Random retry.} If the frame-head attempt fails, the agent stays
silent at the remaining frame positions except $Q_i(t)=R_i(t)$, where it
retries once.
\item \emph{Full-frame failure.} If both attempts fail, the agent remains
silent through the end of the current frame. After position $L-1$, it
starts a new frame in the next slot with $Q_i\leftarrow0$ and a newly drawn
$R_i$.
\item \emph{Success reset.} If any attempt succeeds, then in the next slot
$\Delta_i=1$, $Q_i=-1$, $R_i=0$, and any pending retry is canceled.
\end{enumerate}
After any frame-start initialization, the transmission decision is
\begin{equation}
U_i(t)=\mathbf{1}\{Q_i(t)=0\}+\mathbf{1}\{Q_i(t)=R_i(t)\},
\label{eq:tx-rule}
\end{equation}
where the two terms are mutually exclusive since $R_i(t)\ge1$ whenever the
agent is active. Local frames are asynchronous across agents: a frame starts
when the agent's own age reaches the threshold or its previous frame ends
without success. Neither a common frame
boundary, inter-agent signaling, nor knowledge of a global slot index is
required. Each agent thus needs only its own age and two
local variables and ACK feedback. Thus, AFRA requires synchronization of
slot boundaries, but no alignment of local frames or centralized scheduling.

Fig.~\ref{fig:protocol-timeline} illustrates the evolution of the AoI and local
states for $T=L=5$, including failed attempts, a successful retry, and the
cancellation of a pending retry after a successful frame-head transmission.

\section{Theoretical Analysis}
\label{sec:analysis}

\label{sec:steady}

This section analyzes AFRA's long-run behavior. We first formulate its
Markov state and define a closed set of TDMA-like states with collision-free
periodic operation. We then prove that the system enters this set with
probability one from any legal initial state under $L\ge T>N\ge2$, and
derive its steady-state throughput, average AoI, and average PAoI.

\subsection{Markov State Formulation}
\label{sec:markov-state}
Let $X_i(t)\triangleq(\Delta_i(t),Q_i(t),R_i(t))$ and
$X(t)\triangleq(X_1(t),\dots,X_N(t))$. By the protocol rules, a state is
\emph{legal} if every agent's triple lies in
\begin{equation}
\begin{aligned}
\mathcal X_{T,L}^{\mathrm{node}} \triangleq{}&
\{(\delta,-1,0):1\le\delta\le T-1\}\\
&{}\cup \{(T+kL,0,r)\}
\cup \{(T+q+kL,q,r)\},
\end{aligned}
\label{eq:legal-state}
\end{equation}
where $k\in\mathbb{Z}_{\ge0}$ and $q,r\in\{1,\dots,L-1\}$; the three subsets
correspond to an inactive agent, an agent at its frame head, and an agent inside
its frame, respectively. The system state space is
$\mathcal X_{T,L}\triangleq(\mathcal X_{T,L}^{\mathrm{node}})^N$. Given
$X(t)$, the transmission decisions are determined, the channel outcome follows
from the set of transmitters, and the only new randomness comes from the
independent uniform redraws of $R_i$ at frame starts; hence $\{X(t)\}$ is a
time-homogeneous Markov chain on the countable space $\mathcal X_{T,L}$, with
one-step update
\par
{\small
\begin{equation}
X_i(t+1)=
\begin{cases}
(1,-1,0), & Y_i(t)=1,\\[1pt]
(\Delta_i+1,-1,0), & Q_i=-1,\ \Delta_i<T-1,\\[1pt]
(\Delta_i+1,0,R_i'), & Q_i=-1,\ \Delta_i=T-1,\\[1pt]
(\Delta_i+1,Q_i+1,R_i), & 0\le Q_i\le L-2,\\[1pt]
(\Delta_i+1,0,R_i'), & Q_i=L-1,
\end{cases}
\label{eq:state-update}
\end{equation}
\par
}
\noindent where the first branch corresponds to $Y_i(t)=1$ and the remaining branches
to $Y_i(t)=0$, $R_i'$ is an independent uniform draw on $\{1,\dots,L-1\}$, and
the right sides use the values at slot $t$.

\subsection{TDMA-Like Steady-State Set}
\label{sec:tdma-set}
We now define the states corresponding to collision-free periodic operation
and establish that the system remains in this set once it enters.

\begin{definition}
\label{def:tdma}
The set of \emph{TDMA-like} states is
\begin{equation}
\mathcal S_T \triangleq \bigl\{ x\in\mathcal X_{T,L}:
\begin{array}{l}
1\le\Delta_i\le T,\ \Delta_i\ne\Delta_j\ (i\ne j),\\[1pt]
\Delta_i<T \Rightarrow (Q_i,R_i)=(-1,0),\\[1pt]
\Delta_i=T \Rightarrow Q_i=0
\end{array}
\bigr\}.
\label{eq:tdma-set}
\end{equation}
\end{definition}
\vspace{2pt}

\begin{proposition}
$\mathcal S_T$ is nonempty if and only if $T\ge N$.
\label{prop:feasible}
\end{proposition}

\begin{proof}
The $N$ ages in a TDMA-like state are distinct values in $\{1,\dots,T\}$,
which is possible only if $T\ge N$. Conversely, if $T\ge N$, assign $N$
distinct ages to the agents, keep agents with age below $T$ inactive, and put
the agent with age $T$, if any, at its frame head; the resulting state
satisfies \eqref{eq:tdma-set}.
\end{proof}

The next result shows that $\mathcal S_T$ is absorbing: once the system enters
this set, it remains there and runs collision-free with period $T$.

\begin{lemma}
If the state lies in $\mathcal S_T$, then in the current slot at most one
agent transmits: the agent with age $T$, if present, transmits alone and
succeeds, and otherwise the slot is idle. Moreover, the next state also lies
in $\mathcal S_T$, the ages evolve as $\Delta_i(t+1)=f_T(\Delta_i(t))$ with
$f_T(\delta)=(\delta\bmod T)+1$, and the system runs deterministically with
period $T$ and without collisions.
\label{lem:closure}
\end{lemma}

\begin{proof}
Distinct ages imply that at most one agent has age $T$, and only such an agent
transmits, so the transmission is collision-free. By the update rule
\eqref{eq:state-update}, this agent resets to $(1,-1,0)$, while every other
agent is inactive and ages; an agent whose age reaches $T$ starts a frame at
its head. The map $f_T$ is a cyclic permutation of $\{1,\dots,T\}$, so the ages
stay distinct and in range, and applying it repeatedly shows that the state
repeats every $T$ slots.
\end{proof}

\subsection{Almost-Sure Convergence}
\label{sec:convergence}
We next show that the system reaches $\mathcal S_T$ with probability one
under $L\ge T>N\ge2$. Let
\begin{equation}
\tau_{\mathcal S}\triangleq\inf\{t\ge0: X(t)\in\mathcal S_T\}
\label{eq:tau}
\end{equation}
be the absorption time into $\mathcal S_T$. The next
lemma characterizes $\mathcal S_T$ in terms of the system's recent success
history.

\begin{lemma}
$X(t)\in\mathcal S_T$ if and only if every agent has succeeded at least once
in the last $T$ slots $t-T,\dots,t-1$.
\label{lem:covering}
\end{lemma}

\begin{proof}
Let $s_i(t)$ be the last success slot of agent $i$ before slot $t$, so that
$\Delta_i(t)=t-s_i(t)$. If every agent succeeded within the last $T$ slots,
then $1\le\Delta_i(t)\le T$ and the ages are distinct because a slot delivers
at most one report. An agent with $\Delta_i(t)<T$ has not reached the threshold
again and is inactive; an agent with $\Delta_i(t)=T$ reaches the threshold at
slot $t$ and starts a frame at its head. Hence all conditions in
Definition~\ref{def:tdma} hold, so $X(t)\in\mathcal S_T$. Conversely, if
$X(t)\in\mathcal S_T$, then $1\le\Delta_i(t)\le T$ by
Definition~\ref{def:tdma}, so every agent succeeded within the last $T$ slots.
\end{proof}

By Lemma~\ref{lem:covering}, entry into $\mathcal S_T$ is equivalent to every
agent having succeeded at least once within the preceding $T$ slots. This
characterization underlies the following proof of almost-sure convergence
from any admissible initial state.

\begin{proposition}
\label{prop:path}
Assume $L\ge T>N\ge2$. From every legal initial state $x$, the system enters
$\mathcal S_T$ with probability one, i.e.,
\begin{equation}
\Pr\nolimits_x\{\tau_{\mathcal S}<\infty\}=1.
\label{eq:positive-path}
\end{equation}
\end{proposition}
\vspace{5pt}

\begin{proof}
We establish reachability of the steady-state set from any legal initial
state by constructing a finite path with positive probability, and then
use the finite state space to conclude almost-sure convergence.

First, clear the retries pending in the initial state so that subsequent
transmissions can be constructed without interference from these retries.
Consecutive frame-head transmissions of each agent are at least $T$ slots
apart, so any $T$ consecutive slots contain at most $N<T$ frame-head
transmissions and hence a slot with none. During this clearing step,
prescribe each newly drawn retry position as $R=1$, leaving initially
fixed positions unchanged. After $L$ slots, all initially pending retries have
been executed or canceled. Take a subsequent slot $u$ with no frame-head
transmission. Any remaining retry comes from a frame-head failure in
slot $u-1$ and is executed in slot $u$, in which no new pending retry is created.
Thus, no retries remain pending at the beginning of slot $u+1$.

Next, construct the subsequent retry sequence. Suppose $k$ agents collide at their
frame heads in slot $b$, with $P$ previously scheduled retries still
pending. Maintain the requirement that each prescribed retry occurs
within $T-1$ slots of its collision. In slots $b+1,\dots,b+T-1$,
these $P+k$ agents do not reach another frame head while waiting;
if a retry succeeds, the next frame head also lies beyond this window.
Each remaining agent makes at most one frame-head transmission, leaving
at least $(T-1)-(N-P-k)\ge P+k$ slots without one. Excluding the $P$
scheduled retries leaves $k$ distinct available slots for the newly
collided agents to succeed alone, with $1\le R_i\le T-1\le L-1$.
Apply the same construction at each subsequent frame-head collision.

The construction terminates in finite time. Each agent's next
frame head, as determined when pending retries are cleared, occurs in
finite time; the agent then succeeds either immediately or at its
prescribed retry. Currently periodic agents cannot collide with one
another: each new success avoids their transmissions, and all subsequent
transmissions have period $T$. The remaining processed agents are
waiting for their scheduled retries. Thus, every new frame-head collision
involves an agent that has not yet reached its first such frame head,
so only finitely many collisions occur. Once the remaining retries
finish, all agents operate periodically without collisions. Within one
further period, the covering criterion of Lemma~\ref{lem:covering}
holds, and the system enters $\mathcal S_T$.

This trajectory prescribes finitely many legal retry positions, each
with probability $1/(L-1)>0$, and therefore has positive probability.
For fixed $N,T,L$, replacing each age by $\min\{\Delta_i,T+1\}$ while retaining
frame and retry positions yields a finite access state space.
Since $\mathcal S_T$ is closed and reachable with positive probability
from every outside state, no closed communicating class lies outside it.
A finite-state Markov chain eventually enters a closed recurrent class
with probability one. Hence, $\Pr_x\{\tau_{\mathcal S}<\infty\}=1$.
\end{proof}

\begin{remark}
\label{rem:boundary}
Nonemptiness of $\mathcal S_T$ alone does not guarantee convergence from
every legal initial state. When $T=N$, even if agents occupy distinct
modulo-$T$ transmission slots, an obstacle arises in clearing pending
retries: all $N$ slots within a period are occupied by frame-head
transmissions, leaving no slot free of transmission. Consequently, any
pending retry inevitably collides with some agent's frame head. While that
retry is consumed, the collided agent is forced to schedule a new pending
retry, causing retries to circulate indefinitely among the occupied slots
and preventing the system from clearing all conflicts to enter $\mathcal S_T$.
The configuration $N=T=L=2$ represents the most extreme case where the sole
retry position causes both agents to collide at every slot; for $N>2$, the
same obstacle manifests as a pending retry circulating among agents' frame
heads. Therefore, almost-sure convergence from all legal initial states
strictly requires the margin condition $T>N$ (i.e., $T\ge N+1$), ensuring
at least one slot free of frame heads to safely absorb pending retries.
\end{remark}

\subsection{Steady-State Performance}
\label{sec:steady-performance}
\begin{proposition}
\label{prop:perf}
Under $L\ge T>N\ge2$, the system enters $\mathcal S_T$ almost surely, and the
long-run throughput, average AoI, and average PAoI are
\begin{equation}
\eta=\frac{N}{T},\qquad
\overline{\mathrm{AoI}}=\frac{T+1}{2},\qquad
\overline{\mathrm{PAoI}}=T.
\label{eq:steady-perf}
\end{equation}
\end{proposition}
\vspace{5pt}

\begin{proof}
By the convergence argument above, $\tau_{\mathcal S}<\infty$ almost surely,
so the transient does not affect long-run averages. Inside $\mathcal S_T$,
Lemma~\ref{lem:closure} shows that the ages follow the period-$T$ orbit
$f_T$; each agent visits $1,\dots,T$ once per period and succeeds exactly once,
at age $T$. Hence the system delivers $N$ reports per $T$ slots, each agent's
time-average age is $\frac{1}{T}\sum_{\delta=1}^{T}\delta=(T+1)/2$, and every
delivery occurs at peak age $T$.
\end{proof}

\section{Numerical Results}
We evaluate AFRA through simulations under the system model in
Section~\ref{sec:system}, focusing on convergence and information freshness
over finite observation windows. We compare AFRA with Original TSA,
1-persistent TSA, and SAT, using TDMA as an ideal scheduling reference. These
comparisons examine convergence time, window-averaged AoI, and throughput
across age thresholds, as well as AoI for different network scales.

\subsection{Comparison with Other Access Protocols}

% BEGIN GENERATED PROTOCOL COMPARISON TABLE
\begin{table}[!t]
\centering
\caption{Protocol comparison with $N=100$ and $H=100N$}
\label{tab:protocol-comparison}
\footnotesize
\setlength{\tabcolsep}{2pt}
\renewcommand{\arraystretch}{1.05}
\begin{tabular}{lrrrrrr}
\toprule
Protocol & $T$ & $p$ & $E_{\mathrm{st}}$ & $\eta_H$ & $\overline{\Delta}_{H}$ & $\overline{\Delta}_{\infty}$\\
\midrule
Original TSA & 198 & 3.075\% & -- & 0.360 & 153.84 & 149.63 \\
\midrule
1-pers. TSA & 200 & 2.134\% & 1715 & 0.489 & 107.32 & 100.50 \\
 & 190 & 1.924\% & 2766 & 0.507 & 105.86 & 95.50 \\
 & \textbf{189}$^{\star}$ & 1.894\% & 2641 & 0.510 & \textbf{105.12} & 95.00 \\
 & 180 & 1.560\% & 6655 & 0.512 & 111.23 & 90.50 \\
 & 170 & 1.120\% & 21144 & 0.490 & 135.57 & 85.50 \\
 & 160 & 0.756\% & 159360 & 0.489 & 163.04 & 80.50 \\
 & 150 & 0.525\% & $>10^6$ & 0.484 & 207.79 & 75.50 \\
 & 140 & 0.384\% & $>10^6$ & 0.481 & 268.93 & 70.50 \\
 & 130 & 0.292\% & $>10^6$ & 0.476 & 352.50 & 65.50 \\
 & 120 & 0.229\% & $>10^6$ & 0.476 & 451.89 & 60.50 \\
 & 110 & 0.184\% & $>10^6$ & 0.472 & 569.30 & 55.50 \\
\midrule
AFRA & 200 & -- & 1341 & 0.490 & 109.86 & 100.50 \\
 & 190 & -- & 1424 & 0.515 & 104.93 & 95.50 \\
 & 180 & -- & 1430 & 0.543 & 100.01 & 90.50 \\
 & 170 & -- & 1491 & 0.574 & 95.36 & 85.50 \\
 & 160 & -- & 1482 & 0.609 & 90.27 & 80.50 \\
 & 150 & -- & 1688 & 0.646 & 86.49 & 75.50 \\
 & 140 & -- & 1802 & 0.690 & 82.31 & 70.50 \\
 & 130 & -- & 2117 & 0.737 & 79.58 & 65.50 \\
 & \textbf{122}$^{\star}$ & -- & 2412 & 0.780 & \textbf{77.91} & 61.50 \\
 & 120 & -- & 2564 & 0.790 & 78.55 & 60.50 \\
 & 110 & -- & 3811 & 0.837 & 84.16 & 55.50 \\
\midrule
SAT & 271 & -- & -- & 0.367 & 139.72 & 136.00 \\
\midrule
TDMA & -- & -- & -- & 1.000 & 50.50 & 50.50 \\
\bottomrule
\end{tabular}
\vspace{2pt}
\begin{minipage}{\columnwidth}\scriptsize
$\star$: minimum window AoI in the integer scan; the 1-pers. access probability minimizes the analytical transient time.
$>10^6$: the measured mean exceeds the cutoff, or all 100 runs remain unfinished at the cutoff.
Steady-state AoI is estimated after burn-in for Original TSA and SAT and is analytical otherwise.
Both TDMA AoI entries use $(N+1)/2$; no entry time is reported.
\end{minipage}
\end{table}
% END GENERATED PROTOCOL COMPARISON TABLE
Extensive simulations demonstrate that AFRA achieves the fastest convergence
when =T$. We therefore set =T$ for AFRA in the following comparisons.
Table~\ref{tab:protocol-comparison} compares AFRA with Original TSA~\cite{yavascan2021threshold},
1-persistent TSA~\cite{dester2026onepersistent}, SAT~\cite{chen2022age},
and TDMA for $N=100$ and $H=100N$, using 100 seeds per configuration.
Threshold-based schemes use an unfavorable synchronized initial state:
all agents start at their respective age thresholds and transmit simultaneously.
Here $\eta_H$ is window throughput (successful transmissions per slot),
and $\overline{\Delta}_H$ and $\overline{\Delta}_{\infty}$ are window and steady-state AoI.
For each threshold, the 1-persistent access probability minimizes the
expected transient time of the auxiliary model in~\cite{dester2026onepersistent}.
Original TSA~\cite{yavascan2021threshold} uses the empirically optimal
configuration from our joint search for minimum window-averaged AoI.

AFRA combines rapid convergence with low AoI at small thresholds.
At $T=110$, it reaches steady operation in about $3.8\times10^3$ slots,
whereas none of the 1-persistent runs finishes within $10^6$ slots.
This enables AFRA to exploit smaller thresholds within a finite window.
At its best sampled threshold $T=122$, AFRA achieves a window AoI of
77.91, 25.9\% lower than the best sampled value of 105.12 for
1-persistent TSA. The corresponding window throughput is 0.780,
compared with 0.510 for 1-persistent TSA at its AoI-minimizing sampled
threshold. AFRA also achieves lower window AoI and higher window
throughput than Original TSA and SAT at these operating points.

\begin{figure}[!htbp]
\centering
\includegraphics[width=\columnwidth]{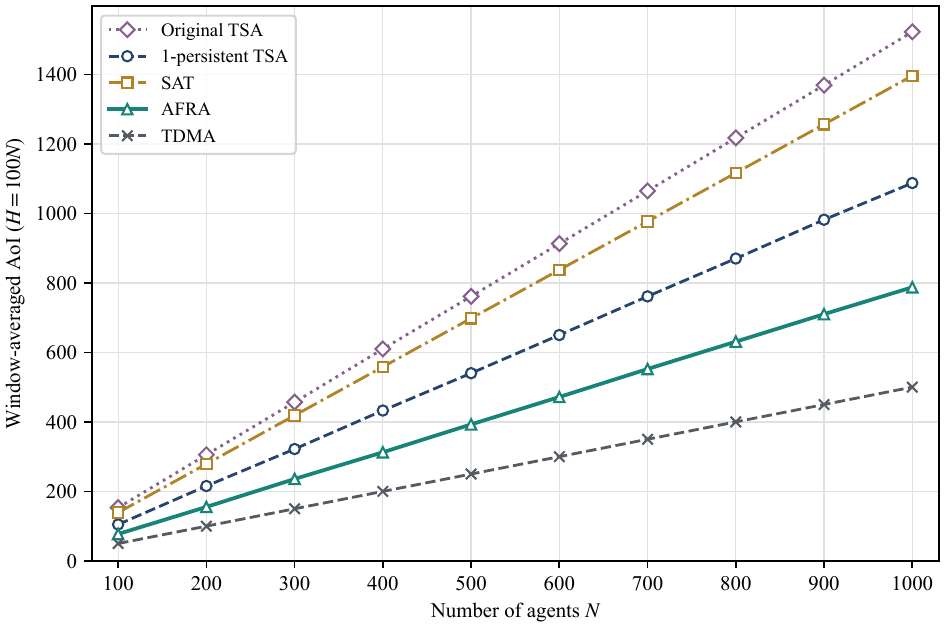}
\caption{Average AoI versus the number of agents $N$ under different
protocols over an observation window of $H=100N$.}
\label{fig:aoi-nodes}
\end{figure}

Fig.~\ref{fig:aoi-nodes} compares different team sizes with $H=100N$.
Original TSA, 1-persistent TSA, and AFRA are evaluated using their
empirically optimal parameters, while SAT adapts its access probability
dynamically.
AFRA reduces window-averaged AoI by 48.5\%, 27.3\%, and 43.7\% relative to
Original TSA, 1-persistent TSA, and SAT, respectively.

\section{Conclusion}
This paper proposed AFRA, an asynchronous frame-based random-access
protocol for multi-agent communication, by modifying TSA to use a
mandatory frame-head transmission and one random retry within each frame.
We proved almost-sure convergence to collision-free steady operation
under $L\ge T>N\ge2$ and characterized its steady-state performance.
For $H=100N$, AFRA
reduces average AoI by 48.5\%, 27.3\%, and 43.7\% relative to Original TSA,
1-persistent TSA, and SAT, respectively.
AFRA also offers a substantial convergence advantage at small thresholds:
for $N=100$ and $T=110$, its mean absorption time is approximately
$3.8\times10^3$ slots, whereas none of the tested 1-persistent TSA runs
reaches the steady-state set within $10^6$ slots.
Future work will analyze the time to reach steady state, optimize thresholds
over finite observation windows, and explore mechanisms that reserve recurring
transmission slots following successful transmissions.

\bibliographystyle{IEEEtran}
\bibliography{references}

@article{yavascan2021threshold,
  author = {Yavascan, Orhan Tahir and Uysal, Elif},
  title = {Analysis of Slotted {ALOHA} With an Age Threshold},
  journal = {IEEE Journal on Selected Areas in Communications},
  volume = {39},
  number = {5},
  pages = {1456--1470},
  year = {2021},
  doi = {10.1109/JSAC.2021.3065043}
}

@article{chen2022age,
  author = {Chen, Xingran and Gatsis, Konstantinos and Hassani, Hamed and Bidokhti, Shirin Saeedi},
  title = {Age of Information in Random Access Channels},
  journal = {IEEE Transactions on Information Theory},
  volume = {68},
  number = {10},
  pages = {6548--6568},
  year = {2022},
  doi = {10.1109/TIT.2022.3180965}
}

@article{dester2026onepersistent,
  author  = {Dester, Pl{\'i}nio S. and Souza, Richard Demo and Cardieri, Paulo},
  title   = {Analysis of the 1-Persistent Age Threshold Slotted {ALOHA}},
  journal = {IEEE Communications Letters},
  volume  = {30},
  pages   = {1066--1070},
  year    = {2026},
  doi     = {10.1109/LCOMM.2026.3662201}
}

@inproceedings{yates2017status,
  author = {Yates, Roy D. and Kaul, Sanjit K.},
  title = {Status Updates over Unreliable Multiaccess Channels},
  booktitle = {2017 IEEE International Symposium on Information Theory (ISIT)},
  pages = {331--335},
  year = {2017},
  doi = {10.1109/ISIT.2017.8006544}
}

@article{maatouk2020csma,
  author = {Maatouk, Ali and Assaad, Mohamad and Ephremides, Anthony},
  title = {On the Age of Information in a {CSMA} Environment},
  journal = {IEEE/ACM Transactions on Networking},
  volume = {28},
  number = {2},
  pages = {818--831},
  year = {2020},
  doi = {10.1109/TNET.2020.2971350}
}

@article{liu2026sequence,
  author = {Liu, Fang and Wong, Wing Shing and Lo, Yuan-Hsun and Zhang, Yijin and Chen, Chung Shue},
  title = {Analysis Methodology for Age of Information Under Sequence-Based Scheduling},
  journal = {IEEE Transactions on Information Theory},
  volume = {72},
  number = {4},
  pages = {2201--2217},
  year = {2026},
  doi = {10.1109/TIT.2026.3667001}
}

@article{tripathi2023fresh,
  author = {Tripathi, Vishrant and Jones, Nicholas and Modiano, Eytan},
  title = {{Fresh-CSMA}: A Distributed Protocol for Minimizing Age of Information},
  journal = {Journal of Communications and Networks},
  volume = {25},
  number = {5},
  pages = {556--569},
  year = {2023},
  doi = {10.23919/JCN.2023.000032}
}

@article{ahmetoglu2022mista,
  author = {Ahmetoglu, Mutlu and Yavascan, Orhan Tahir and Uysal, Elif},
  title = {{MiSTA}: An Age-Optimized Slotted {ALOHA} Protocol},
  journal = {IEEE Internet of Things Journal},
  volume = {9},
  number = {17},
  pages = {15484--15496},
  year = {2022},
  doi = {10.1109/JIOT.2022.3179132}
}

@article{munari2021modern,
  author = {Munari, Andrea},
  title = {Modern Random Access: An Age of Information Perspective on Irregular Repetition Slotted {ALOHA}},
  journal = {IEEE Transactions on Communications},
  volume = {69},
  number = {6},
  pages = {3572--3585},
  year = {2021},
  doi = {10.1109/TCOMM.2021.3060429}
}

@article{yue2023frame,
  author = {Yue, Zhiling and Yang, Howard H. and Zhang, Meng and Pappas, Nikolaos},
  title = {Age of Information Under Frame Slotted {ALOHA}-Based Status Updating Protocol},
  journal = {IEEE Journal on Selected Areas in Communications},
  volume = {41},
  number = {7},
  pages = {2071--2089},
  year = {2023},
  doi = {10.1109/JSAC.2023.3280990}
}

@inproceedings{shi2021communication,
  author = {Shi, Guangyao and Rabban, Ishat E. and Zhou, Lifeng and Tokekar, Pratap},
  title = {Communication-Aware Multi-Robot Coordination with Submodular Maximization},
  booktitle = {2021 IEEE International Conference on Robotics and Automation (ICRA)},
  pages = {8955--8961},
  year = {2021},
  doi = {10.1109/ICRA48506.2021.9561041}
}

@article{guo2018gathering,
  author = {Guo, Meng and Zavlanos, Michael M.},
  title = {Multirobot Data Gathering Under Buffer Constraints and Intermittent Communication},
  journal = {IEEE Transactions on Robotics},
  volume = {34},
  number = {4},
  pages = {1082--1097},
  year = {2018},
  doi = {10.1109/TRO.2018.2830370}
}

@inproceedings{tian2024ihero,
  author = {Tian, Zhuoli and Zhang, Yuyang and Wei, Jinsheng and Guo, Meng},
  title = {{iHERO}: Interactive Human-Oriented Exploration and Supervision Under Scarce Communication},
  booktitle = {Proceedings of Robotics: Science and Systems},
  year = {2024},
  doi = {10.15607/RSS.2024.XX.115}
}

@inproceedings{pongsirijinda2023gmc,
  author = {Pongsirijinda, Khattiya and Cao, Zhiqiang and Shalihan, Muhammad and Ng, Benny Kai Kiat and Lau, Billy Pik Lik and Yuen, Chau and Tan, U-Xuan},
  title = {{GMC-Pos}: Graph-Based Multi-Robot Coverage Positioning Method},
  booktitle = {2023 IEEE International Conference on Robotics and Biomimetics (ROBIO)},
  pages = {1--7},
  year = {2023},
  doi = {10.1109/ROBIO58561.2023.10354990}
}

\end{document}